# Influence of the voltage taps position on the self-field DC and AC transport characterization of HTS superconducting tapes

M. Vojenčiak[a,*], F. Grilli[a], A. Stenvall[b], A. Kling[a], W. Goldacker[a]

[a]Karlsruhe Institute of Technology, Institute for Technical Physics, Hermann-von-Helmholtz-Platz 1, 76344 Eggenstein-Leopoldshafen, Germany
[b]Electromagnetics, Tampere University of Technology, P.O. Box 692, FI-33101 Tampere, Finland

*Corresponding author at: Karlsruhe Institute of Technology, Institute for Technical Physics, Hermann-von-Helmholtz-Platz 1, 76344 Eggenstein-Leopoldshafen, Germany.
Tel.: +49 721 608-23531
E-mail address: michal.vojenciak@kit.edu

**Abstract**

The current-voltage (I-V) curve is the basic characteristic of a superconducting wire or tape. Measuring I-V curves is generally problematic when samples have poor stabilization. Soldering voltage taps to an active part of the conductor affects the effectiveness of the local cooling and/or can be difficult to do in certain devices such as fault current limiters and cables where the tapes are closely packed. In order to overcome these problems, voltage taps can be placed outside the active area of the superconductor. We proved both by simulations and experiments that this arrangement leads to the same results as the standard four point method and it provides more detailed information for sample protection. The same arrangement can also be used for AC transport loss measurement. However in this case particular care has to be taken because the eddy current loss in the current leads contributes to the total measured loss. We used numerical simulations to evaluate the contribution of the eddy current loss to the measured AC loss. With help of simulations one can determine whether the contribution of the eddy current loss is significant and possibly optimize the current leads to reduce that loss contribution.

## 1. Introduction

The current-voltage (I-V) curve is the basic characteristic of a superconducting wire or tape. Its shape and dependence on various parameters, such as applied magnetic field, temperature, and mechanical stress, give important information on the superconductor itself as well as on the whole composite conductor [1, 2, 3]. In the case of more complex assemblies (coils, cables), it determines the performance of the whole superconducting device.

During the measurement of the I-V curve the applied current reaches the critical current of the superconductor. Dissipative processes such as flux creep and flux flow take place at such high current [2, 4, 5, 3]. At a certain level of dissipated power, the sample reaches its thermal runaway point and subsequent processes (transition of the superconductor to the normal state, change of cooling regime etc.) can damage the sample [6]. In case of high-temperature superconductors (HTS) a partial quench damage leading to stepwise decrease of the critical current has been observed [7]. In general, the measurement of the I-V curve is problematic in cases of samples with poor stabilization at given conditions (temperature, magnetic field). In the vicinity of the current leads, the transport current flows in the normal metal [8]; additional heat produced in the resistive material causes the thermal runaway to start in that area. In a standard four-point measurement, voltage taps are placed at a certain distance from the current leads. In such an arrangement the critical part of the superconductor is not monitored. Especially at high operation temperatures, the quench propagates very slowly. The quench may not reach the measurement zone before the conductor is damaged.

Soldering voltage taps to the active part of the conductor affects local thermal conditions and cooling effectiveness. A high temperature under a voltage tap during quench can lead to local damage of the superconductor [9]. This may cause problems in case of samples where homogeneous quench is required, e.g. resistive fault current limiters [10].

The experimental problems described above led us to investigate advantages and drawbacks of an alternative voltage taps arrangement – voltage taps placed outside the current leads. In this paper, we investigate the influence of the voltage taps position on the correctness of the measurements, their precision and the protection of the sample. We also use numerical simulations to verify the experimental results.

This arrangement can also be used for AC transport loss measurements, where it provides the advantage of sample protection and of monitoring the whole superconductor carrying current. The only difference with respect to the measurement of the I-V characteristics is that the wiring should be placed at a certain distance from the conductor in the case of a flat tape [11]. However, the loop formed by the signal wires measures also the loss induced by the magnetic field in the superconductor's ambient. Since current leads are usually massive blocks of copper with high conductivity, an eddy current loss in the current leads can be even higher than the loss in the superconductor itself. We use numerical simulation to determine the contribution of the eddy current loss with respect to the measured loss.

In the present work we focus on HTS coated conductors; however similar analysis can be made also for other kinds of superconductors and we believe that results will be similar.

In this paper we first describe the arrangement of the voltage taps and the experimental setup – section 2. For validating our experimental observations, both in AC and in DC conditions, we use numerical simulations. The calculation methods are described in section 3 – FEM simulation. Then we continue by discussing DC I-V curves in section 4 – DC current-voltage characteristic. Section 5 deals with transport AC loss. In this section we discuss experimental results and show how to correctly interpret the experimental results with the help of numerical simulations. Finally we summarize our observations in section 6 – conclusions.

**2. Voltage taps arrangement and experimental setup**

In this paper we investigate three different arrangements of the voltage taps, as schematically shown in figure 1. The lead wire to the voltage taps is placed in plane with coated conductor in distance about 12 mm from edge of the conductor (see also figure 2) . Voltage taps D-F are usually used for measurements on short samples or simple devices. Contacts C-G are usually employed for characterization of more complex devices such as cables or fault current limiters. The disadvantage of using contacts C-G is that the measured voltage consists of two components: the voltage along the superconductor and the voltage on contact resistances. In order to avoid measuring this latter voltage and at same time not to place contacts on the active part of the conductor we placed voltage taps outside the current leads – contacts A and I in figure 1. Such voltage taps are usually used for quench detectors.

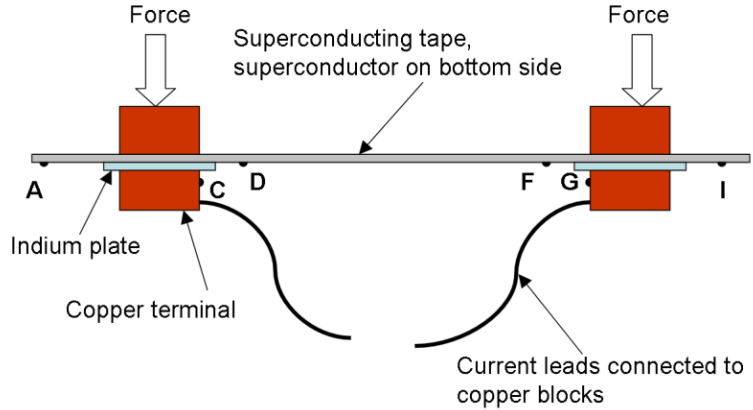

**Figure 1.** Position of the various pairs of voltage taps: A-I, C-G, D-F.

Measurements were performed on commercial *RE*BCO (*RE* = rare earth) coated conductor prepared by IBAD technology on hastelloy substrate [12]. The width of the tape was 12 mm and its self-field critical current was about 300 A. The total length of the sample was 110 mm, the distance between current leads was 55 mm and the width of the current leads was 12 mm. All the measurements were performed at 77 K.

The sample was connected to the current leads by pressed contacts. There was an indium foil placed between the copper current leads and the silver cap layer of the superconducting tape. The voltage taps were soldered by indium and we used them for AC as well as for DC measurements. The wiring between the voltage taps was placed about 20 mm outside the tape edge for correct measurement of the transport AC loss. The distance between voltage taps D-F was 30 mm. A detailed view of the sample holder is shown in figure 2 a.

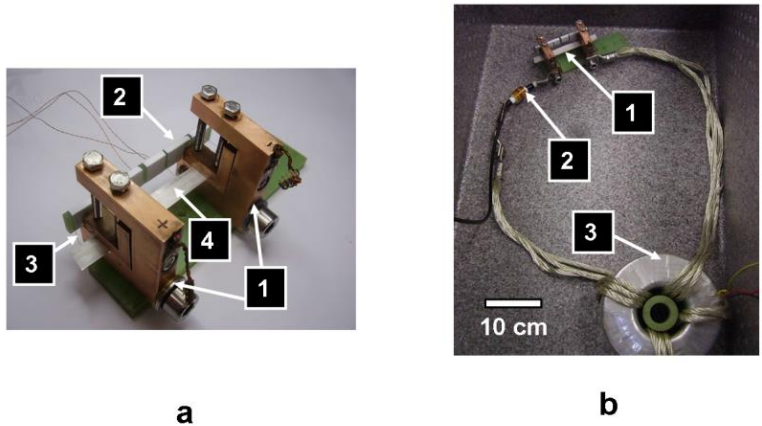

**Figure 2.** a – Detailed view of the sample holder with current leads and voltage taps: 1 – current leads, 2 – support for wiring of measuring loops, 3 – wiring of voltage tap A, 4 – sample. b – Experimental setup for AC loss measurements: 1 – sample, 2 – Rogowski coil, 3 – transformer.

For feeding the AC current we used a toroidal transformer immersed in $LN_2$ bath together with the sample. For AC loss measurement we used a phase-sensitive detector (lock-in amplifier) [13]. The amplitude of the current and the reference phase was provided by the signal from the Rogowski coil – see figure 2 b.

### 3. Finite element method simulations

Three dimensional finite element method (FEM) simulations were performed for two purposes. First, to check the value of the resitivitity $\rho$ of the copper used in the sample holder as current leads. Second, to obtain information about eddy currents induced in the current leads from the time varying current in the sample.

For evaluating the resistivity of the current leads, we modelled the actual shape of the current lead and fed current from the real current contact to the sample– see figure 3 a. The current output was in the tape connection to the current lead. This was implemented in the model by setting electric ground to that surface. We solved for the stationary current problem

$$\nabla \cdot \frac{1}{\rho} \nabla \varphi = 0 \qquad (1)$$

where $\varphi$ is the electric potential. This approach was selected since in the simulations we were interested only in the behaviour of the current lead. In the measurement, test currents varying from -20 to 20 A were supplied to the measurement system and voltages between voltage taps $V_0$ and $V_1$ shown in figure 3 a were measured. After a linear fit procedure, a value for resistance was determined (0.34 µΩ). In the simulations, the copper resistivity was adjusted so that it provided the same resistance as the measured one - see figure 3 b. A simulated potential distribution is shown in figure 3 a. (The picture shows normalized values, since the potential distribution qualitatively looks always the same for the different values of current). For the resistivity we obtained 2.11 nΩm at 77 K, which agrees with the tabulated values of 1.9 – 2.3 nΩm [1].

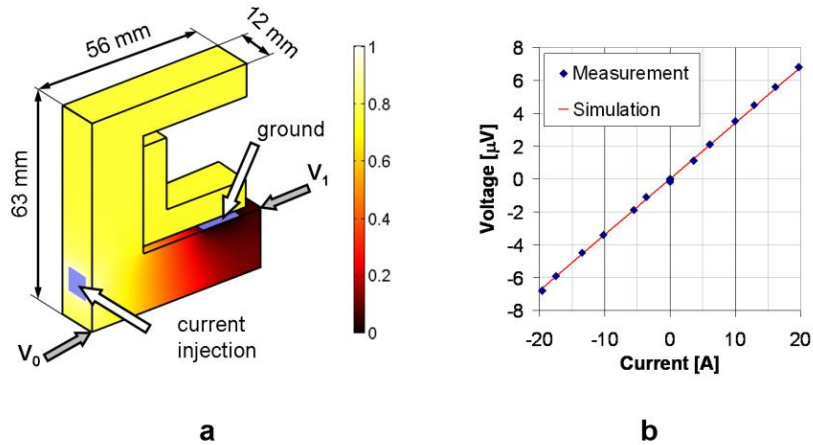

**Figure 3.** a - distribution of electric potential (normalized to unity) in the sample holder b – comparison of measured and calculated voltage between voltage taps $V_0$ and $V_1$ shown in figure a, simulation used $\rho$ = 2.11 nΩm

In order to study the eddy currents induced in the current lead by the current flowing in the sample, it is necessary not to have any driving current in the current lead. Otherwise it would not be possible to separate the losses due to the induced current and the losses due to the transport current since there is also skin effect with the transport current. Thus, in the simulations, we placed the copper block in the middle of the 11 cm long tape and forced a homogeneous given net current $J_e$ into the tape. The principal difference between the experiment and the modelled geometry is schematically shown in figure 4. The region of the copper blocks where most of the eddy current flows and most of the dissipation occurs is shown by the shaded areas.

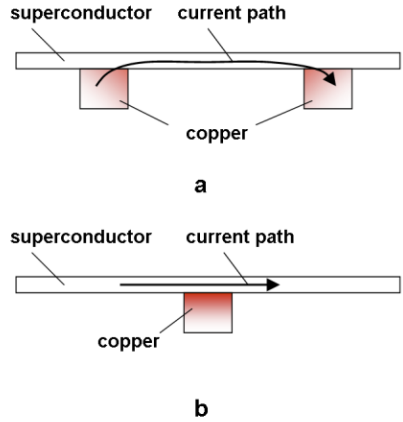

**Figure 4.** Sketch of geometry for eddy-current loss simulation. Shading shows qualitatively distribution of eddy current loss in copper current leads. a – situation in experiment, b – situation in model

A time harmonic eddy current problem formulated in terms of magnetic vector potential $A$ and electric potential $\varphi$

$$j2\pi f \sigma A + \sigma \nabla \varphi + \nabla \times \frac{1}{\mu} \nabla \times A = J_e \qquad (2)$$

where $f$ is the frequency, $\sigma$ is the conductivity and $\mu$ is the permeability, was then solved in the modelling domain. It was necessary to use $\sigma$ instead of $\rho$ since in the air region $\rho$ diverges toward an infinite value, whereas 0 for $\sigma$ can be used. The value $\sigma = 0$ was used for the tape as well. In this way, the current distribution in the tape could be imposed to be homogeneous. From the solution, the losses in the sample holder were determined by integrating the average ohmic power density. From this value, the loss per cycle was determined.

**4. DC current-voltage characteristic**

To investigate properties of a superconductor one needs to measure an electric field distribution along the superconductor. The voltage taps are soldered to a stabilizing resistive material (not directly to the superconductor) and thus the measured electric field can generally differs to one along the superconductor.

First we investigated the current and the electric field distribution in the parts of the sample close to the current leads by means of numerical simulations. The electric current flows from the current lead into the superconductor through a stabilization layer or a metal matrix, in case of a coated conductor and a wire respectively. For a certain distance from the current lead part of the current flows in the normal metal stabilization and/or metal matrix. The ratio between the current flowing in the normal metal and that flowing in the superconductor depends on the distance from the current lead and it is characterized by the so-called transfer length [1, 14, 15]. It was experimentally shown that this length is well below 1 mm for samples similar to ours [15]. To see details of the potential distribution along the length of the sample we made numerical simulations.

For calculating the electric potential distribution we used a similar model to that shown in figure 3, but the point with ground potential was shifted to the middle of the superconductor at a distance of 27 mm from the current leads. The superconductor was modelled as a material with conductivity $\sigma = 10^{15}$ S/m. Figure 5 shows the calculated electric potential distribution on top of the silver cap layer in the part of the superconductor close to the current lead.

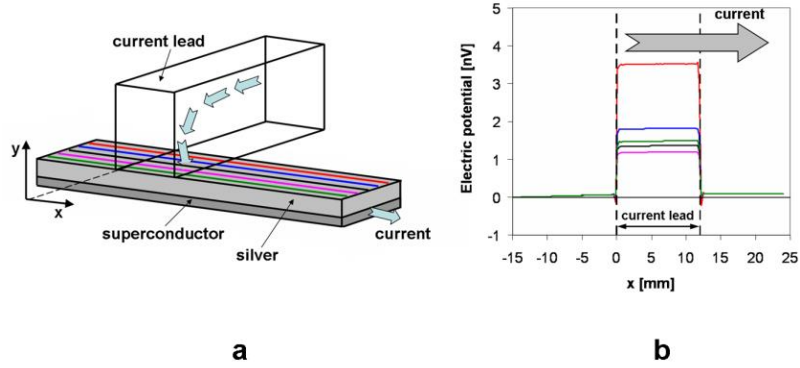

**Figure 5**. Distribution of the electric potential on the surface of the silver cap layer; the driving current is 100 A. a – lines where electric potential was calculated b – distribution of electric potential along the lines shown in figure a

An electric potential under the current lead (0 < x < 12 mm) is necessary for current transfer through the silver cap layer. From the results one can see that the voltage on the silver depends on the position along the width of the sample too. Along the sample length, in the direction toward the centre of the sample (positive x-axis), the electric potential drops to a very low value and reaches 0 in the middle of the sample. In the direction outside the current leads, where the voltage taps A-I are placed (negative x-axis) the electric potential also drops down to a very low value. This proves that the voltages taps A-I are almost at the same potential as the superconductor.

Another important issue for I-V curve measurement is estimation of the distance between the voltage taps. This distance is used for calculation of the electric field and in this way for the critical current evaluation. The distance estimation is straightforward only for the voltage taps D-F. The distance between these contacts can be measured with a negligible error. For the contacts C-G usually the distance between the current leads is used as length between the voltage taps, even though part of the current flows in a superconductor under the current lead already. The physical distance between contacts A-I can be much larger than the distance between the current leads. The current flows from a current lead toward another current lead – see figure 1 and figure 5. Part of the superconductor on the outer side of the current lead does not carry electric current (except extremely small current to a voltmeter). There is no electric field along the superconductor carrying current lower than the critical one, therefore the voltage taps outside the current leads measure only voltage on the part of the superconductor carrying overcritical current, i.e. on the part between the current leads. Therefore one should use the distance between inner edges of the current leads for the electric field estimation.

Current-voltage characteristics measured by various pairs of voltage taps are shown in figure 6 a. The voltage measured by the contacts C-G increases linearly at low currents. One can estimate the sum of resistances between the voltage taps C-G and superconductor from the linear part of the I-V curve. In our experiment the estimated sum of two contact resistances is 0.289 μΩ.

For quantitative analysis of the measured I-V curve we used the fitting equation with two parameters – $I_C$ and $n$:

$$E = E_C \left( \frac{I}{I_C} \right)^n \qquad (3)$$

We used criterion of electric field $E_C = 10^{-4}$ V/m (1 μV/cm).

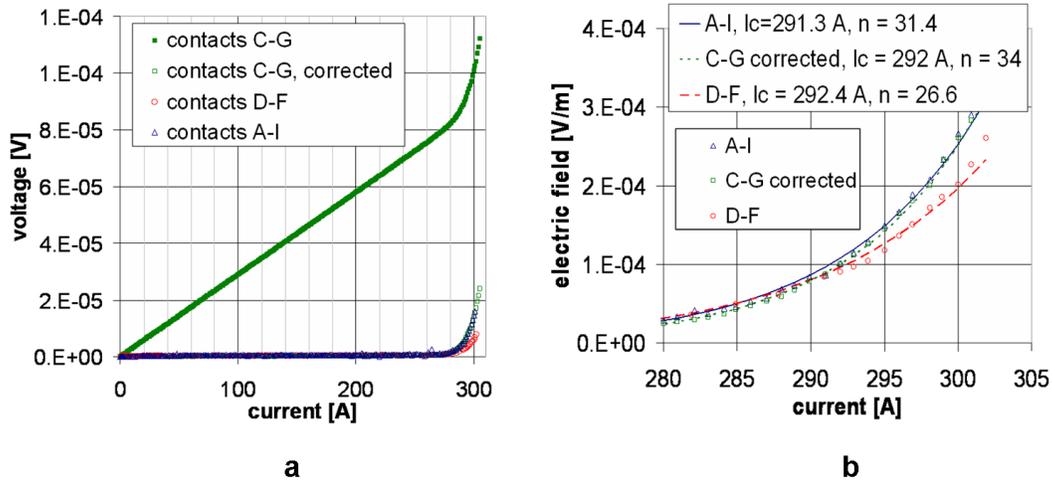

**Figure 6.** a – current-voltage characteristics measured by different pairs of voltage taps. The I-V curve measured between contacts C-G was corrected by subtraction of the resistive voltage, with a resistance value of 0.289 µΩ. b – detail of experimental I-V curves together with fit curves – see equation (3)

The resistive voltage measured by the contacts C-G at the critical current (fig. 6 a) is already more than 80 µV, while the voltage criterion for the critical current estimation is 5.5 µV.

All three measured curves result in a critical current of 292 A. Differences between the estimated critical currents is less than 1 A. The I-V curves measured by the contacts A-I and the contacts C-G are slightly steeper (higher *n*) than the I-V curve measured by the contacts D-F – see figure 6 b. This could be caused by the higher temperature of the superconductor in the vicinity of the current leads. Higher temperature is a consequence of the additional heat produced in the current leads and reduced cooling since the superconductor is not cooled directly by the nitrogen bath. An increase of the *n*-value due to heat produced in the samples was reported also in [16].

The heat produced in the current leads and stabilization can lead to an increase of the superconductor's temperature and to quench. The voltage measured by the voltage taps placed in the central part of the superconductor, where the current flows only in the superconductor, contains no information about the part of the superconductor close to the current leads. Therefore quench detection utilizing the voltage taps placed on the central part of the superconductor can be inefficient.

**5. Transport AC loss**

For the transport AC loss measurement we used the same contacts as for DC current-voltage characterization. Position of the lead wire to the voltage taps strongly influences measured loss in superconductor [17, 18]. In our case we used the lead wire at a distance of about three times the half-width of the tape from the centre of the conductor. The dependence of the transport AC loss on transport current amplitude is shown in figure 7 for the different contacts at various frequencies.

Normalization by the length in figure 7 was performed in a similar way to the DC measurement; the loss measured by contacts A-I and contacts C-G was normalized by the distance between the current leads (55 mm), while the loss measured by contacts D-F was normalized by the real distance between the voltage taps (30 mm). While in case of DC I-V curve a small error in estimation of the distance between contacts A-I does not change the value of the critical current significantly, in the case of AC transport current the situation is different. Also an applied current lower than the critical one causes dissipation. The physical distance between inner edges of the current leads (55 mm) and distance between current leads plus two times the width of the current leads (79 mm) constitute the lowest and highest limit for the distance estimation. This difference is significant for short samples. In the following, we used 55 mm.

There are 3 groups of curves shown in figure 7. Curves measured by contacts D-F are lines without symbols, curves measured by contacts A-I are marked by triangles and diamonds; curves measured by contacts C-G are marked by stars. Norris's analytical prediction for a thin superconducting strip transporting AC current is shown by a thick black line [19].

Standard transport AC loss measurement is represented by measurement utilizing the contacts D-F. The loss measured by these contacts agrees well with the analytical prediction and is basically independent of frequency.

The loss measured on contacts C-G is strongly dependent on the frequency. This is expected, because the loss in the contact resistance (0.289 µΩ) is included in this voltage. The resistive loss is dominant and therefore the loss per cycle decreases with increasing frequency.

Also the loss measured on contacts A-I has a rather strong frequency dependence. Measurements in a wide range of frequencies show a non-monotonic dependence of the loss per cycle on frequency. The range of the measured frequencies was from 2.25 Hz up to 360 Hz. We extracted the dependence of the loss measured by the contacts A-I on frequency for a fixed amplitude of current 50 A – figure 8. In this figure we used values of loss not normalized by the length, because we analyze also the loss in the current leads, which does not depend on the length of the superconductor between them.

There are also two calculated curves shown in figure 8. The black dashed line shows the loss in the superconductor calculated by Norris's analytical model [19]. This loss is frequency-independent, since Norris's calculations are based on the critical state model [20]. The loss caused by eddy currents in the copper current leads was calculated with our FEM model. The calculated eddy current loss in the copper current leads agrees quite well with the measured loss; the largest difference is around 10%. This agreement suggests that the used measuring loop (formed by voltage taps and lead wiring to them) picks up most of the signal caused by eddy-currents (however, we did not investigate influence of the measuring loop size and position on measured loss in details). A small difference is expected, because the real shape of the sample holder and its properties always slightly differ from the situation in the FEM model. From comparison of the three curves in figure 8 it is clear that the eddy currents loss in the copper current leads dominates. In other words, the transport AC loss measured by the contacts outside the current leads includes also the eddy current loss in the current leads. Since current leads are usually massive, one has to take this effect into account.

Although the results of the eddy current loss calculation agree very well with the experimental results, it is not possible to estimate the real value of the loss in superconductor. This later is expected to be around 40 times lower than the loss in the peak of the measured curve (frequency 36 Hz). However, using numerical modelling one can determine the necessary length of the tape between the contacts so that the induced loss is negligible and/or optimize the shape of the current leads. This can be important for measurements on fault current limiters where an alternative positioning of the voltage taps becomes necessary and the tape is usually quite long.

The voltage taps placed outside the current leads are beneficial in case of quench. The current flowing from the current lead to the superconductor has to go through the resistive stabilization layer therefore additional heat is produced in this region. At a certain level of current the local temperature increase causes thermal runaway and the sample quenches locally. The quenched region is usually close to the current lead and thus quench cannot be detected by contacts placed in the standard position D-F. The transport loss measured until the sample quenches is shown in figure 9. The sample quenched at the current amplitude of around 360 A. In this case, the hot-spot appeared close to the current lead (and outside the D-F contacts) and the resulting voltage increase led to limitation of the transport current to around 65 A. Part of the superconductor between contacts D-F was not affected by quench and therefore the current reduction results in decrease of the voltage measured on them. The contacts A-I measure the total voltage along the whole superconductor; including quenched part. Therefore the voltage measured on them increases after quench. A properly adjusted quench detector connected to contacts A-I can prevent an excessive increase of the local temperature and thus the thermal damage of the superconductor.

## 6. Conclusions

Voltage contacts placed directly on the superconducting tape, but outside the current leads, can be beneficial for DC measurements especially for short samples, because they measure the electric field along the superconductor, but do not touch the active part. Such contacts are beneficial for protection of the sample against thermal runaway originating from the heat produced in an area close to the current leads.

Contacts outside the current leads can be used also for transport AC loss measurement. However, in the case of AC transport current, there is also an eddy current loss in the current leads, which is included in the transport loss measured by these contacts.

Since the current leads are usually massive copper blocks, the eddy current loss in the normal metal can dominate the loss in the superconductor. For eddy current loss estimation numerical calculations can be used and, depending on the ratio between the calculated loss in the current leads and the loss in the superconductor, one can decide whether using voltage taps placed outside the current leads is applicable for a particular experiment.

**Acknowledgement**
This work was supported partly by a Helmholtz University Young Investigator Group Grant (VH-NG-617), partly by EFDA under contract WP11-FRF-KIT/Vojenciak, and partly by Academy of Finland, project number 131577/Stenvall.
The authors would like to thank Andrej Kudymow and Bernd Ringsdorf (KIT) for their assistance with experiments.